\newcommand{\be}[0]{\begin{equation}}
\newcommand{\ee}[0]{\end{equation}}
\newcommand{\ba}[0]{\begin{eqnarray}}
\newcommand{\ea}[0]{\end{eqnarray}}

\def\NPB{{\em Nucl. Phys.}}
\def\PLB{{\em Phys. Lett.}}
\def\PRL{\em Phys. Rev. Lett.}

\def\PRD{{\em Phys. Rev.}}

\def\MPLA{{\em Mod. Phys. Lett.} }
\def\EPJ{{\em Eur. Phys. J.}}

\def\etal{{\it et al.}}
\documentclass{ws-ijmpa}
\usepackage{epsfig,multicol}

\begin{document}

\markboth{A. Mirjalili, S. Atashbar Tehrani, Ali N. Khorramian}
{The role of polarized  valons in the flavor symmetry breaking of
nucleon sea}

\catchline{}{}{}{}{}

\title{THE ROLE OF POLARIZED VALONS IN THE FLAVOR SYMMETRY BREAKING OF
NUCLEON SEA}

\author{A. Mirjalili}

\address{Physics Department, Yazd University, Yazd, Iran\\
Institute for Studies in Theoretical Physics and Mathematics
(IPM)\\P.O.Box 19395-5531, Tehran, Iran\\
Mirjalili@ipm.ir}

\author{S. Atashbar Tehrani}

\address{
 Physics Department, Persian Gulf University, Boushehr,
Iran\\
Institute for Studies in Theoretical Physics and Mathematics
(IPM)\\P.O.Box 19395-5531, Tehran, Iran\\
Atashbar@ipm.ir}

\author{\footnotesize Ali N. Khorramian}

\address{Physics Department, Semnan University, Semnan, Iran\\
Institute for Studies in Theoretical Physics and Mathematics
(IPM)\\P.O.Box 19395-5531, Tehran, Iran\\
Alinaghi.Khorramian@cern.ch}

\maketitle

\pub{Received (Day Month Year)}{Revised (Day Month Year)}

\begin{abstract}
Next-to-leading order approximation of the quark helicity
distributions
 are used in the frame work of polarized valon model.
The flavor-asymmetry in the light-quark sea of the nucleon can be
obtained from the contributions of  unbroken sea quark
distributions. We employ the polarized valon model and extract the
flavor-broken light sea distributions which are modeled with the
help of a Pauli-blocking ansatz. Using this ansatz, we can obtain
broken polarized valon distributions. From there and by employing
convolution integral, broken sea quark distributions are
obtainable in this frame work. 0ur results for $\delta u$, $\delta
d$, $\delta \bar u$, $\delta \bar d$ and $\delta \bar s$ are in
good agreement with recent experimental data for polarized parton
distribution from HERMES experimental group and also with GRSV
model. Some information on orbital angular momentum as a main
ingredient of total nucleon spin are given. The $Q^2$ evolution of
this quantity, using the polarized valon model is investigated.

\end{abstract}

\ccode{PACS Nos.: 13.60.Hb, 12.39.-x, 13.88.+e.}
\section{Introduction}
In contrast to most deep inelastic structure functions which are
correspond to spin average scattering, it will be possible to
extend the discussion to the situation where, for instance, the
lepton beam and nucleon target are polarized in the longitudinal
direction. The polarized parton distributions of the nucleon have
been intensively studied in recent years
\cite{ref1}$^-$\cite{ref17}. The conclusion has been that the
experimental data dictate a negatively polarized
anti-quark component, and show a tendency toward a positive polarization of gluons.\\

Presently there are a lot of precise data
\cite{ref18}$^-$\cite{ref27} on the polarized structure functions
of the nucleon. Recently HERMES experimental group has reported
some data \cite{ref28} for the quark helicity distributions in the
nucleon for up, down, and strange quarks from semi-inclusive
deep-inelastic scattering. Among experimental groups, HERMES as a
second generation experiment  can be used to study the spin
structure of the nucleon by measuring not only inclusive but also
semi-inclusive and exclusive processes in deep-inelastic lepton
scattering. Semi-inclusive deep-inelastic is a powerful tool to
determine the separate contributions of polarized distribution of
$\delta q{(x)}$ of the quarks and anti quarks of flavor $f$ to the
total spin of the nucleon. From analytical point of view, all
polarized parton distributions including the polarized light
symmetric sea distribution can be obtained in this letter through
the polarized valon model \cite{ref17}.\\

 Hwa \cite{ref29-1} found
evidence for the valons in the deep inelastic neutrino scattering
data, suggested their existence and applied it to a variety of
phenomena. Hwa \cite{ref29-2} has also successfully formulated a
treatment of the low-$p_{T}$ reactions based on a structural
analysis of the valons. Here a valon can be defined as a valence
quark and associated sea quarks and gluons which arise in the
dressing processes of QCD\cite{ref29-3}. In a bound state problem
these processes are virtual and a good approximation for the
problem is to consider a valon as an integral unit whose internal
structure cannot be resolved. In a scattering situation, on the
other hand, the virtual partons inside a valon can be excited and
be put on the mass shell. It is therefore more appropriate to
think of a valon as a cluster of partons with some momentum
distributions. The proton, for example, has three valons which
interact with each other in a way that is characterized by the
valon wave function, while they  respond independently in an
inclusive hard collision with a $Q^{2}$ dependence that can be
calculated in QCD at high $Q^{2}$. Hwa and Yang \cite{ref29}
refined the idea of the valon
model and extracted new results for the valon distributions.\\

Flavor-broken light sea distributions, using the pauli-blocking
ansatz \cite{ref30} is investigated in this article. As it was
suggested \cite{ref31}, the asymmetry is related to the Pauli
exclusion principle (`Pauli blocking') . Since the  symmetric
polarized sea quark distributions in polarized valon frame work
are obtainable from  \cite{ref17}, we use them and employ the same
technique as in \cite{ref15} to move to broken scenario in which
it is assumed $\delta \bar {u}(x,
Q^2)\neq\delta \bar {d}(x, Q^2)\neq\delta \bar {s}(x, Q^2)$.\\

On the other hand, the measurement of the polarized structure
function $g_1^p(x,Q^2)$ by the European Muon Collaboration (EMC)
in 1988 \cite{ref19} has revealed more profound structure of the
proton, that is often referred to as ‘the proton spin crisis’ in
which the origin of the nucleon spin is one of the  hot problems
in nucleon structure. In particular, it still remains a mystery
how spin is shared among valence quarks, sea quarks, gluons and
orbital momentum of nucleon constituents. The results are
interpreted as very small quark contribution to the nucleon spin.
Then, the rest has to be carried by the gluon spin and/or by the
angular momenta of quarks and gluons. The consequence from the
measurement was that the strange quark is negatively polarized,
which was not anticipated in a naive quark model. Our calculations
which have been done in polarized valon frame work, have resulted
 a negative polarized for strange quark and in agreement with recent data.
 Having the contributions of all  sea quark distributions (broken or unbroken)
 in the singlet sector of distributions, and using the gluon contribution and
helicity sum rule, it will be possible to investigate the total
angular momentum $L_{z}$ which we can attribute to constituent
quark and gluon in a hadron. The $Q^2$ dependence of parton
angular momentum \cite{ref32} is investigated and can be obtained
through the polarized valon model which will be discussed in more details.  \\

This paper is planed as in following. In Sec. 2 we are reviewing
how to extract the polarized parton distributions, using valon
model. Sec. 3 is advocated to calculate polarized light-quark sea
in two unbroken and broken scenarios. We break there the polarized
valon distributions and obtain breaking function in terms of the
unbroken distributions. In Sec. 4, we calculate the first moment
of polarized parton distribution in broken scenario. We discuss
the total angular momentum of quarks and gluons as an important
ingredient in considering the total spin of nucleon in Sec. 5. Its
$Q^2$ evolution is considered in the LO approximation and compered
with the NLO result, using directly helicity sum rule in the valon
model. Sec. 6 contains our conclusions.

\section{Valon model and polarized parton distributions}
To describe the quark distribution $q(x)$ in the valon model, one
can try to relate the polarized quark distribution functions
$q^\uparrow$ or $q^\downarrow$ to the corresponding valon
distributions $G^\uparrow$ and $G^\downarrow$. The polarized valon
can still have the valence and sea quarks that are polarized in
various directions, so long as the net polarization is that of the
valon. When we have only one distribution $q(x,Q^2)$ to analyze,
it is sensible to use the convolution in the valon model to
describe the proton structure in terms of the valons. In the case
that we have two quantities, unpolarized and polarized
distributions, there is a choice of which linear combination
exhibits more physical content. Therefore, in our calculations we
assume a linear combination of $G^\uparrow$ and $G^\downarrow$ to
determine respectively the unpolarized ($G$) and polarized
($\delta G$) valon
distributions. \\

Polarized valon distributions in the next-to-leading approximation
were calculated, using improved valon model \cite{ref17}.
According to the improved valon model, the polarized parton
distribution is related to the polarized valon distribution. On
the other hand, the polarized parton distribution of a hadron is
obtained by convolution of two distributions: the polarized valon
distributions in the proton and the polarized parton distributions
for each valon, i.e.
 \begin{equation}
 \label{eq:conv1}
 \delta
q_{i/p}(x,Q^{2})= \sum_{j}\int_{x}^{1}  \delta
q_{i/j}(\frac{x}{y},Q^{2})\delta G_{j/p}(y)\frac{dy}{y}\;,
\end{equation}
where the summation is over the three valons. Here  $\delta
G_{j/p}(y)$ indicates the probability for the $j$-valon to have
momentum fraction $y$ in the proton. $\delta q_{i/p}(x,Q^{2})$ and
$\delta q_{i/j}(\frac{x}{y},Q^{2})$ are respectively polarized
$i$-parton distribution in the proton and $j$-valon . As we can
see the polarized quark distribution can be related to polarized
valon distribution.\\

Using Eq.~(\ref{eq:conv1}) we can obtain polarized parton
distributions in the proton at different value of $Q^2$. In Fig.
(1)  we have presented the polarized  parton distributions in a
proton at
  $Q^{2}=3\;GeV^{2}$. These distributions have been calculated in the NLO
 approximation and compared  with some theoretical
models \cite{ref14}$^-$\cite{ref17}.

\begin{figure}[tbh]
\centerline{\includegraphics[width=0.7\textwidth]{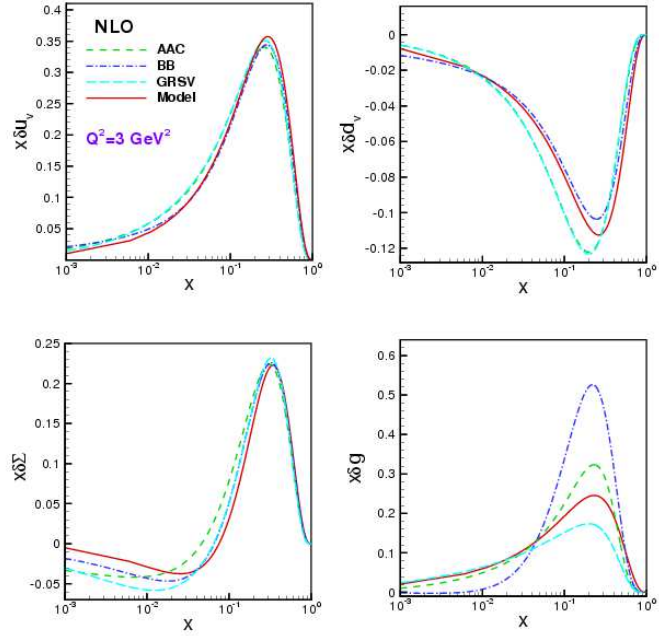}}
\caption{Polarized parton distributions in the proton at $Q^2=3\;
GeV^2$ as a function of $x$ in the NLO approximation. Dashed line
is the AAC model (ISET=3)[14], long-dashed line is the GRSV model
(ISET=1)[15], dashed-dotted line is the BB model (ISET=3)[16] and
the solid line is from [17].}
\end{figure}

\section{Polarized light quark sea in the nucleon}

In spite of many attempts to consider the flavor-asymmetry of the
light quark sea \cite{ref30}, we are interesting to determine the
helicity densities for the up and down quarks and anit-up,
anti-down, and strange sea quarks in the NLO approximation, using
the polarized valon model. We would like to briefly comment on the
assumptions about the polarized anti quark asymmetry made in the
recent analysis of the HERMES data for semi-inclusive DIS
\cite{ref28}.\\
In the following, Subsec. 3.1 , 3.2 are advocated to unbroken  and
broken light quark sea, using the improved valon model.

\subsection{Unbroken scenario}
In this scenario one assumes, as in most analysis of polarization
data performed thus far, a flavor symmetric sea, i.e.
\begin{equation}\label{eq:FlavorSy} \delta \bar {u}(x, Q^2)=\delta  \bar {d}(x,
Q^2)=\delta  \bar {s}(x, Q^2)\equiv\delta  \bar {q}(x, Q^2)\;,
\end{equation} where as usual $\delta u_{sea}=\delta \bar {u}$,
$\delta d_{sea}=\delta \bar {d}$, and $\delta s=\delta \bar {s}$.
The adopted LO and NLO distributions can be taken from the recent
analysis in \cite{ref17}. We have the symmetric nucleon sea
\begin{equation}
\label{eq:SeaSy} \delta \bar {q}(x,Q^2)=\frac{\delta
\Sigma(x,Q^2)-\delta u_v(x,Q^2)-\delta d_v(x,Q^2)}{2f}\;,
\end{equation}
in which the $\Sigma$ symbol denotes $\sum_{q=u,d,s}(q+\bar {q})$.
In Fig. (2)  we have presented the flavor symmetric sea quark
densities, $x\delta \bar q$, as a function of $x$ at $Q^2=1, 2.5,
5, 10, 20, 50 \;GeV^{2}$ values. These distributions were
calculated in the LO and NLO
 approximation, using polarized valon model.
\begin{figure}[tbh]
\centerline{\includegraphics[width=0.5\textwidth]{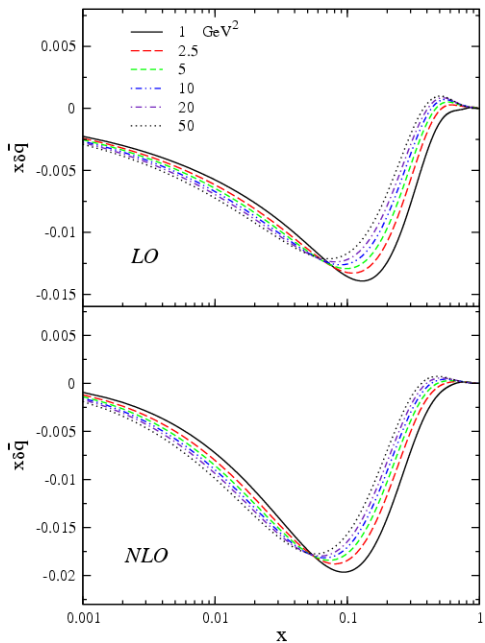}}
\caption{The flavor symmetric sea quark densities in the LO and
NLO approximation as a function of $x$ and for some different
values of $Q^2$.}
\end{figure}\\

By using polarized sea quark distribution and having polarized
valence quark distribution, we can obtain the contributions of
$\delta q(x,Q^2)=\delta q_v(x,Q^2)+\delta \bar q(x,Q^2)$. In Fig.
(3), we used the polarized valon model \cite{ref17} and presented
the $u$ and $d$ helicity quark distributions, $x\delta{u( x,
Q^2)}$ and $x\delta{d( x, Q^2)}$ at $Q^2=2.5 Gev^2$ as a function
of $x$. This result was also
 compared with some other theoretical models in the flavor symmetric
 case [14-16].\\

\begin{figure}[th]
\vspace*{8pt} \centerline{\psfig{file=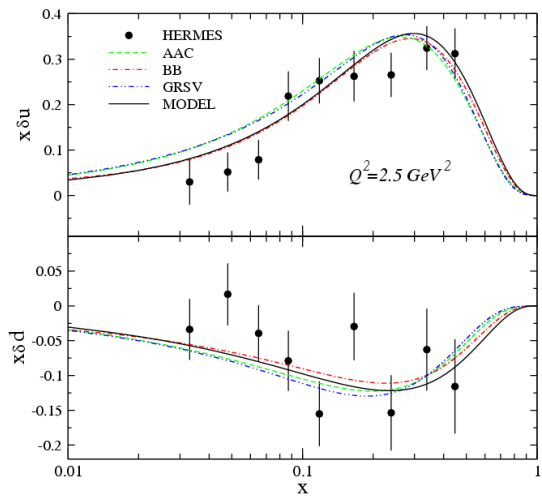,width=2.8in}}
\vspace*{8pt} \caption{The $u$ and $d$ quark helicity
distributions, $x\delta u(x,Q^2)$ and $x\delta d(x,Q^2)$,
evaluated at a common value of $Q^2=2.5 GeV^2$ as a function of
$x$ in the  flavor symmetric case. The dashed line is the AAC
model (ISET=3)[14], dashed-dotted-dotted line is the GRSV model
(ISET=1)[15], dashed-dotted line is the BB model (ISET=3)[16]  and
the solid line is from [17].}
\end{figure}

\subsection{Broken scenario}
We  assume here $\delta \bar {u}(x, Q^2)\neq\delta \bar {d}(x,
Q^2)\neq\delta \bar {s}(x, Q^2)$, i.e. a broken flavor symmetry as
motivated by the situation in the corresponding unpolarized case
\cite{ref33}. Present polarization data provide detailed and
reliable information concerning flavor symmetry breaking and
therefore there is enough motivation to study and utilize
antiquark distributions extracted
via the phenomenological ansatz.\\

To study the asymmetric nucleon sea we refer to an input scale
$Q^2_0=1\; GeV^2$. In this situation we can use
 the Pauli-blocking relation \cite{ref34}
 for the unpolarized and polarized antiquark distributions as

\begin{equation}
\label{eq:PaulAnz1} \frac {\bar {d}(x,Q^2_0)}{\bar
{u}(x,Q^2_0)}=\frac { {u}(x,Q^2_0)}{
 {d}(x,Q^2_0)}\;,
\end{equation}
and \begin{equation}
 \label{eq:PaulAnz2}
 \frac {\delta \bar
{d}(x,Q^2_0)}{\delta \bar {u}(x,Q^2_0)}=\frac {\delta
{u}(x,Q^2_0)}{\delta
 {d}(x,Q^2_0)}\;.
\end{equation}

These equations are essential to study the flavor asymmetry of the
unpolarized and polarized light sea quark distributions which
imply that $u>d$ and consequently will determine $\bar u<\bar d$
and so on. This is in accordance with the suggestion of Feynman
and Field This is correspond to the suggestion of Feynman and
Field \cite{ref31} that, since there are more $u$- than
$d$--quarks in the proton, $u\bar{u}$ pairs in the sea are
suppressed more than $d\bar{d}$ pairs by the exclusion principle.
These two  relations require obviously the idealized situation of
maximal Pauli-blocking and hold approximately in some effective
field theoretic  models \cite{ref30}.\\

The flavor-broken light sea distributions are modeled  in perfect
and interesting way  by Gl\"uck and \etal$\;$  in \cite{ref35}. In
that paper it was trying to use pauli-blocking ansatz and the
introduction a breaking function at $Q^2=Q^2_0$, the valance and
sea quark distributions were improved. Through this model, the
necessary information about the broken symmetry of sea quark
distributions has been
obtained.\\
\subsection{Asymmetric polarized valon distributions}
Here we are trying to use the same strategy as in subsection 3.2
to break the polarized valon distributions which calculated in
\cite{ref17}. Let us begin from the definition of polarized valon
distribution functions
 \ba
\left\{
\begin{array}{ll}
\delta G^{}_{j}(y)=\delta {\cal {W}}^{}_{j}(y)\times G_{j}(y)\  & {\rm for\;non-singlet\;case\;,} \\
\delta G^{'}_{j}(y)=\delta {\cal {W}}^{'}_{j}(y)\times G_{j}(y)\ &
{\rm for\; singlet\; case}\;,
\end{array}
 \right.
\ea where \begin{equation} \delta {\cal
{W}}^{}_{j}(y)=N_{j}y^{\alpha _{j}}(1-y)^{\beta _{j}}(1+\gamma
_{j}y+\eta _{j}y^{0.5})\;,
\end{equation}
the subscript $j$ refers to $U$ and $D$-valons. The motivation for
choosing this functional form is that the low-$y$ behavior of the
valon densities is controlled by the term $y^{ \alpha_j}$ while
that at the large-$y$ values is controlled by the term
$(1-y)^{\beta_j}$. The remaining polynomial factor accounts for
the additional medium-$y$ values. For $\delta {\cal
{W}}^{'}_{j}(y)$ in Eq.(6) we choose the following form
\begin{equation}
\delta {\cal {W}}^{'}_{j}(y)=\delta {\cal {W}}^{}_{j}(y)\times
\sum_ {m=0}^{5} {\cal A}_{m}y^{\frac {m-1}{2}}\;.
\end{equation}\\
The extra term in the above equation, ($\sum$ term), serves to
control the behavior of the singlet sector at very low-$y$ values
in such a way that we can extract the sea quark contributions.
Moreover, the functional form for $\delta {\cal {W}}^{}$ and
$\delta {\cal {W}}^{'}$ give us the best fitting $\chi^2$
value\cite{ref17}.

Consequently we can obtain all flavor asymmetric quark
distributions in the valon model frame work. The flavor-asymmetric
and flavor-symmetric
 valon distributions are denoted  by
$\delta \widetilde{G}$ and $\delta G$ respectively. Since the
valon distributions play the role of quark distributions at low
value of $Q^2$, so we can define a breaking function as in
\cite{ref35} to determine broken polarized valon distributions
from unbroken ones. These distributions are related to each other
as

\[
2\delta \widetilde{G}_{U}(y)\equiv 2\delta G_{U}(y)-\Theta (y)\;,
\]
\begin{equation}
\label{eq:SeaAsym}
 \delta \widetilde{G}_{D}(y)\equiv \delta
G_{D}(y)+\Theta (y)\;,
\end{equation}

here $\Theta$ is called `Breaking' function and the factor 2
indicates the existence of two $U$-valons. As we will see, to
determine this function, we need to obtain the contributions of
sea quarks in the improved valon model framework.

 By using \cite{ref17} we can get the following expressions for
the polarized parton distributions in a proton:

\ba \delta u_v(x,Q^2)&=&2\int_x^1  \delta
f^{NS}(\frac{x}{y},Q^2)\;\delta G^{}_{U}(y)\;\frac{dy}{y}\;,\nonumber \\
\delta d_v(x,Q^2)&=&\int_x^1  \delta
f^{NS}(\frac{x}{y},Q^2)\;\delta G^{}_{D}(y)\;\frac{dy}{y}\;,\nonumber \\
\delta \Sigma(x,Q^2)&=&\int_x^1  \delta
f^{S}(\frac{x}{y},Q^2)\left(2\;[\delta G^{'}_{U}(y)+
\delta G^{'}_{D}(y)]\right)\;\frac{dy}{y}\;,\nonumber \\
\ea here $\delta f^{NS}$ and $\delta f^{S}$ indicate the
Non-singlet and Singlet parton distributions inside the valon. To
obtain the $z$-dependence of parton distributions, $\delta
f^{NS,\;S}(z=\frac{x}{y},Q^2)$, from the $n-$%
dependent exact analytical solutions in the Mellin-moment space,
one has to perform a numerical integral in order to invert the
Mellin-transformation \cite{ref17}.\\

 At the input scale $Q_0^2=1\; GeV^2$,
$\delta f^{NS}$ and $\delta f^{S}$ behave like  the delta function
and consequently $\delta u_v(x,Q_0^2)$ is approaching to $2\delta
G^{}_{U}$ and similarly $\delta d_v(x,Q_0^2)$ to $\delta
G^{}_{D}$. Finally $\delta \Sigma(x,Q_0^2)$ will be equal $2\delta
G_{U}'(y)+\delta G_{D}'(y)$. We should notice that in this scale,
$x$-space is equal to $y$-space. Now  the first term in the
numerator of Eq.~(\ref{eq:SeaSy}) is equal to $2\delta
G_{U}'(y)+\delta G_{D}'(y)$ and the rest two terms to $2\delta
G_{U}(y)+\delta G_{D}(y)$, therefor  the Eq. (3) at $Q^2=Q_0^2$
can be expressed as a combination of $\delta G_{j}$ and $\delta
G_{j}'$ in the following form
\begin{equation} \label{eq:delqbary}
\delta G^{\bar q}\equiv\frac{1}{2f}\sum_j (\delta G_{j}'- \delta
G_{j}),
\end{equation}
where $j$ denotes the $U, U, D$-valons. Here $\delta G^{\bar q}$
denotes to polarized see quark contribution at $Q^2=Q_0^2$ which
is arising out from symmetric valon distributions. \\
In order to determine breaking function, we need the following
relations
\[
\delta G^{\bar u}(y)\equiv \delta G^{\bar q}%
(y)+\frac{\Theta (y)}{2}\;,
\]
\begin{equation}\label{eq:dbar/ubar3}
\delta G^{\bar d}(y)\equiv \delta G^{\bar q}%
(y)-\frac{\Theta (y)}{2}\;,
\end{equation}
where $\delta G^{\bar u}$ and $\delta G^{\bar d}$ indicate
polarized see quark contributions which are obtained from
asymmetric valon
distributions.\\

 In the valon model framework,
Eq.~(\ref{eq:PaulAnz2}) can be considered as
\begin{equation}\label{eq:dbar/ubar1}
 \frac {\delta G^{\bar d}(y)}{\delta G^{\bar u}(y)}=\frac {2\delta \widetilde{G}_U(y)+{\delta G^{\bar u}(y)}}{\delta
\widetilde{G}_D(y)+{\delta G^{\bar d}(y)}}\;.
\end{equation}
On the other hand from  Eq.~(\ref{eq:dbar/ubar3}) we have
\begin{equation}\label{eq:dbar/ubar2}
\frac {\delta G^{\bar d}(y)}{\delta G^{\bar u}(y)}=\frac {\delta
G^{\bar q}(y)-\frac{\Theta(y)}{2}}{\delta G^{\bar
q}(y)+\frac{\Theta(y)}{2}}\;.
\end{equation}
So by using Eqs.~(\ref{eq:dbar/ubar1},\ref{eq:dbar/ubar2}) and
inserting the related functions from Eq.~(\ref{eq:SeaAsym}) we
arrive at

\begin{equation}
\label{eq:equalAnz} \frac {\delta G^{\bar
q}-\frac{\Theta}{2}}{\delta G^{\bar q}+\frac{\Theta}{2}}=\frac
{2\delta {G}_U-\Theta+\delta G^{\bar q}+\frac{\Theta}{2}}{\delta
{G}_D+\Theta+\delta G^{\bar q}-\frac{\Theta}{2}}\;.
\end{equation}
The breaking function can be extracted from
Eq.~(\ref{eq:equalAnz}) as follows
\begin{equation}
\label{eq:theta} \Theta \equiv -2\delta G^{\bar q}\;\frac{2\delta
G_U-\delta G_D}{2\delta G_U+\delta G_D}\;.
\end{equation}
It is obvious  that the  combination of  Eqs.(9,11) will lead to
 the following constrains
\[
2\delta \widetilde{G}_U(y)+2\delta G^{\bar u}(y)=2\delta
{G}_U(y)+2\delta G^{\bar q}(y)\;,
\]
\[
\delta \widetilde{G}_D(y)+2\delta G^{\bar d}(y)=\delta
{G}_D(y)+2\delta G^{\bar q}(y)\;,
\]
\begin{equation}
\label{eq:constrain1} 2\delta \widetilde{G}_U(y)+\delta
\widetilde{G}_D(y)=2\delta {G}_U(y)+\delta {G}_D(y)\;,
\end{equation}
in which, for instance, the first equation in Eq. (17) has been
obtained from the combining of first equation in Eq. (10) with the
first equation  in Eq. (12). Using these constrains, it will be
seen that the first moment of $g_1^p$ will not  change in the
broken sea scenario.\\

Since  the polarized valon  and sea quark distributions are known
in  \cite{ref17}, by substituting them in Eq.~(\ref{eq:theta}),
the breaking function $\Theta$  can be simply parameterized in the
LO and NLO approximations as
\begin{equation}
\label{eq:thetaLO} y\Theta_{LO}(y)
=0.060y^{0.496}(1-y)^{7.735}(1+10.443y+3.371y^{2}-5.759y^{3})\;,
\end{equation}
\begin{equation}
 \label{eq:thetaNLO}
y\Theta_{NLO}(y)=0.192y^{0.619}(1-y)^{7.228}(1-0.037y-0.310y^{2}+1.273y^{3})\;,
\end{equation}
which needed for
performing the $Q^2$-evolution in the $x$-space with using the convolution integral.\\

If we back to Eq.~(\ref{eq:conv1}) and substitute the related
broken valon distributions, we will be able to obtain the $Q^2$
dependent of the parton distribution in broken scenario. Our
results for the polarized parton distributions at $Q^2=5\; GeV^2$
are presented in Fig. (4). In this figure a comparison between the
distributions in the LO (broken scenario) and NLO approximations
(broken and unbroken scenario) has been done.

 It was assumed that $\delta s=\delta
{\overline{s}}\equiv \delta \overline{q}$. We also did not
consider any asymmetry for the strange quark and gluon
distributions. In Fig. (5), the quark helicity distributions
$x\delta q(x,Q^2)$ for $q=u,d,\bar u,\bar d$ and $s$ are shown
 at value of $Q^2=2.5\;GeV^2$ in the
LO and NLO approximation. We have also compared in this figure our
model with recent HERMES \cite{ref28} data and GRSV
model\cite{ref15}.
\begin{figure}[tbh]
\centerline{\includegraphics[width=0.8\textwidth]{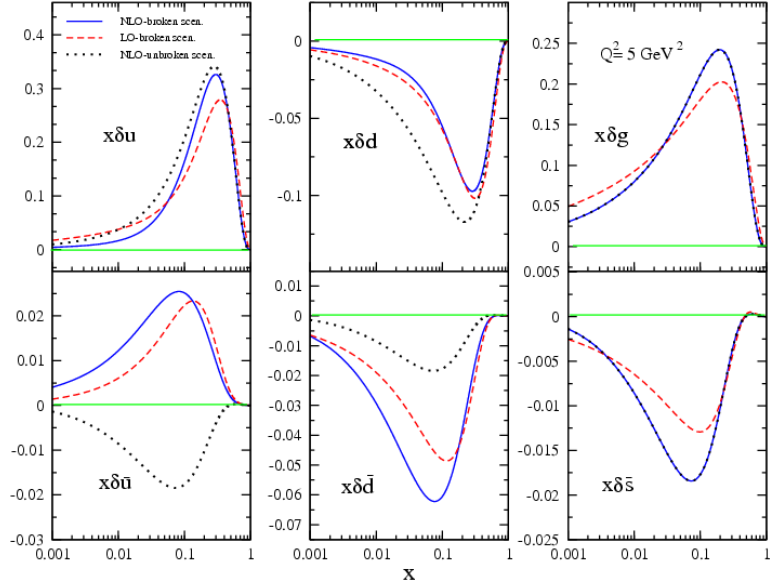}}
\caption{The LO (broken scenario) and NLO approximations (broken
and unbroken scenario), for polarized parton distributions at
$Q^2=5\;GeV^2$. The line is the  NLO distributions in broken
scenario , the dotted line is the NLO distributions in unbroken
scenario and the dashed line is the LO distributions in broken
scenario.}
\end{figure}
\begin{figure}[ht]
\includegraphics[width=.49\linewidth,clip]{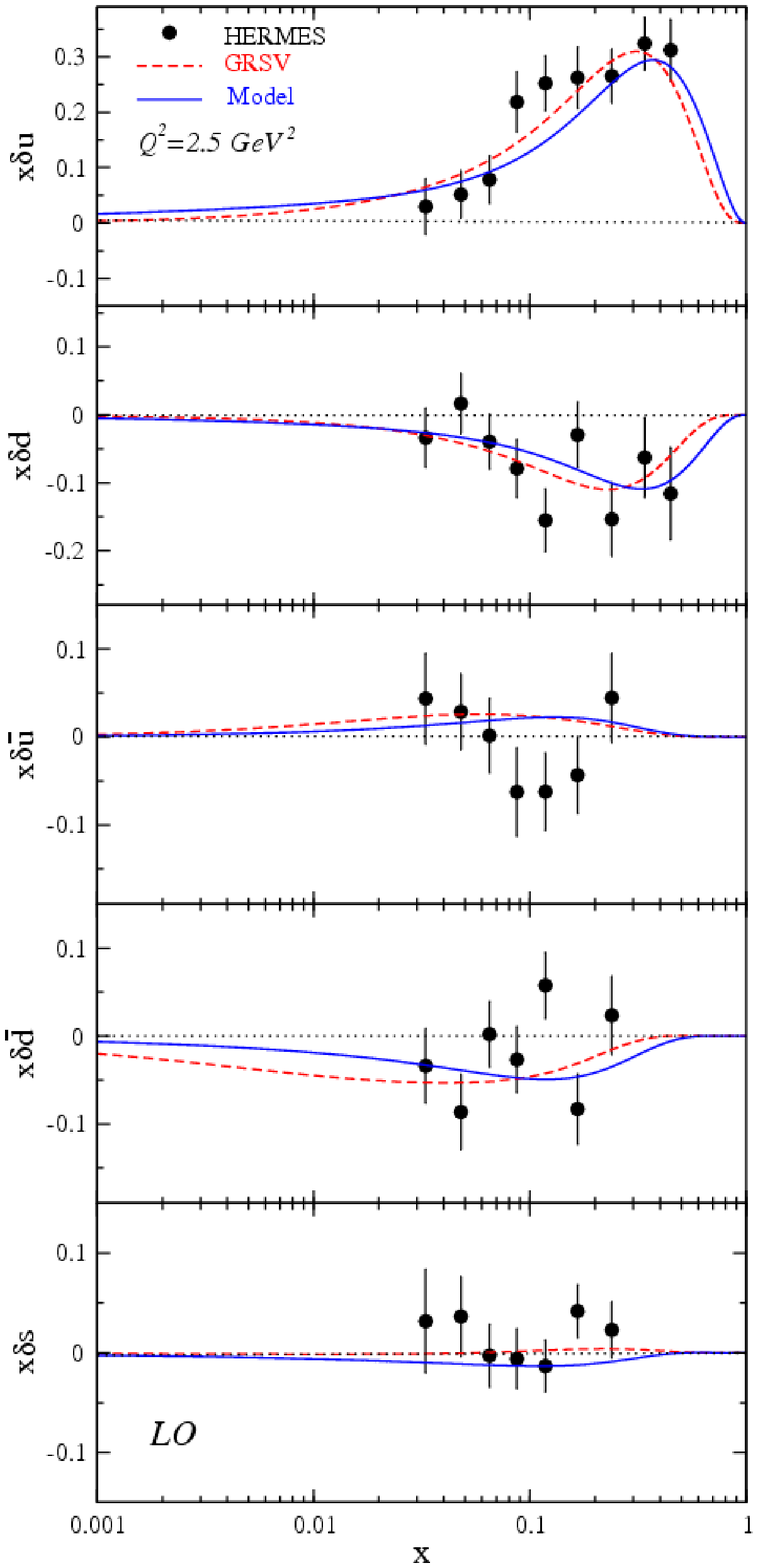}
\includegraphics[width=.49\linewidth,clip]{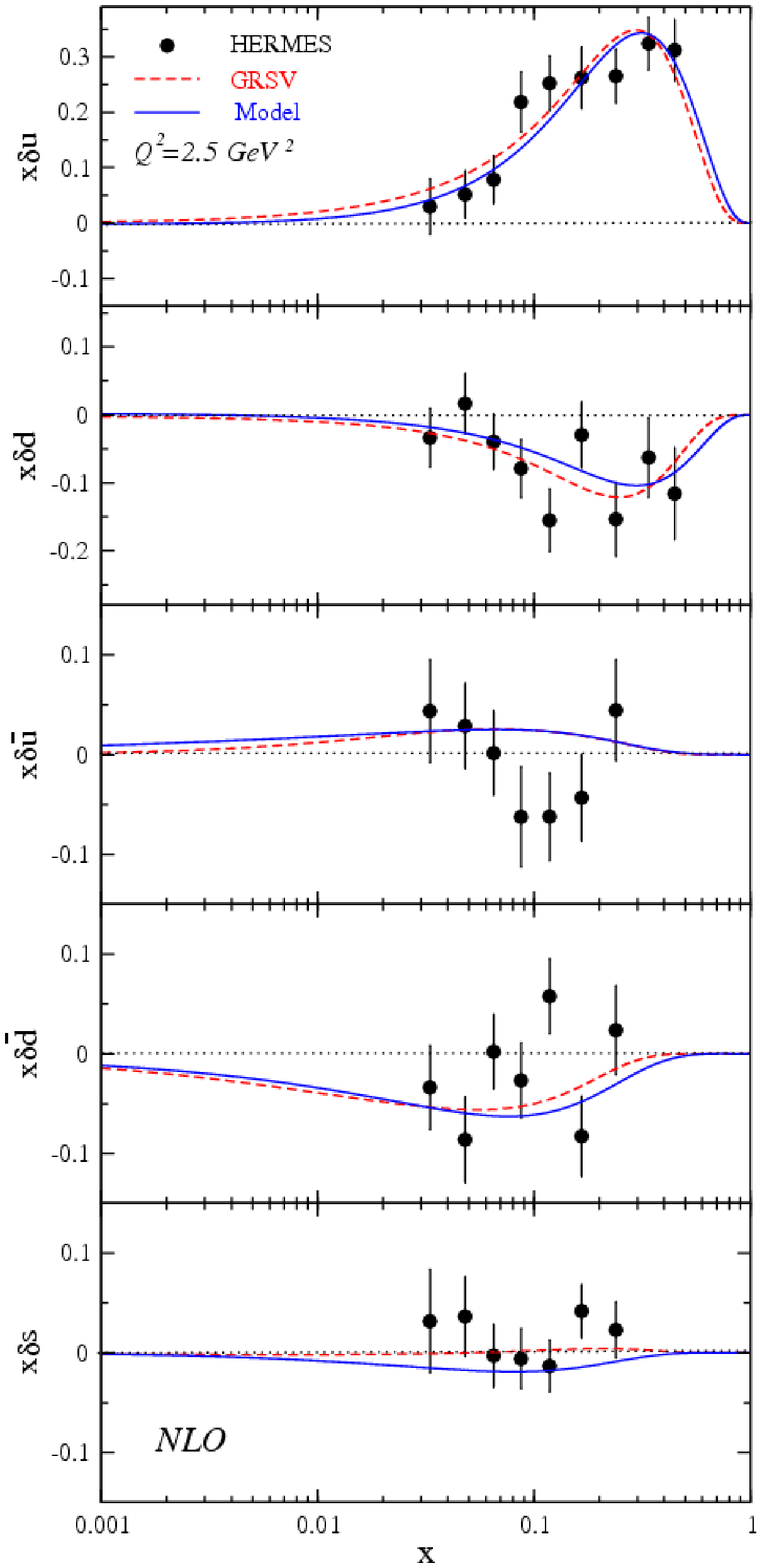}
\vskip-.25cm \caption{The quark helicity distributions $x\delta
q(x,Q^2)$ evaluated at value of $Q^2=2.5\;GeV^2$ as a function of
$x$ in the LO and NLO approximations. The dashed line is the GRSV
model [15](ISET=2 for the NLO and ISET=4 for the LO approximation)
and the solid line is our model.} \vspace*{-4pt}
\end{figure}
\\
After calculating $x\delta \bar u$ and $x\delta \bar d$ for
instance at $Q^2=2.5\; GeV^2$, we can obtain the results of
$x\delta \bar u-x\delta \bar d$. In Fig. (6), we have presented
this function in the broken scenario evaluated in the LO, NLO
approximations and compared it with the recent HERMES \cite{ref28}
data and GRSV model\cite{ref15}.

\begin{figure}[tbh]
\centerline{\includegraphics[width=0.8\textwidth]{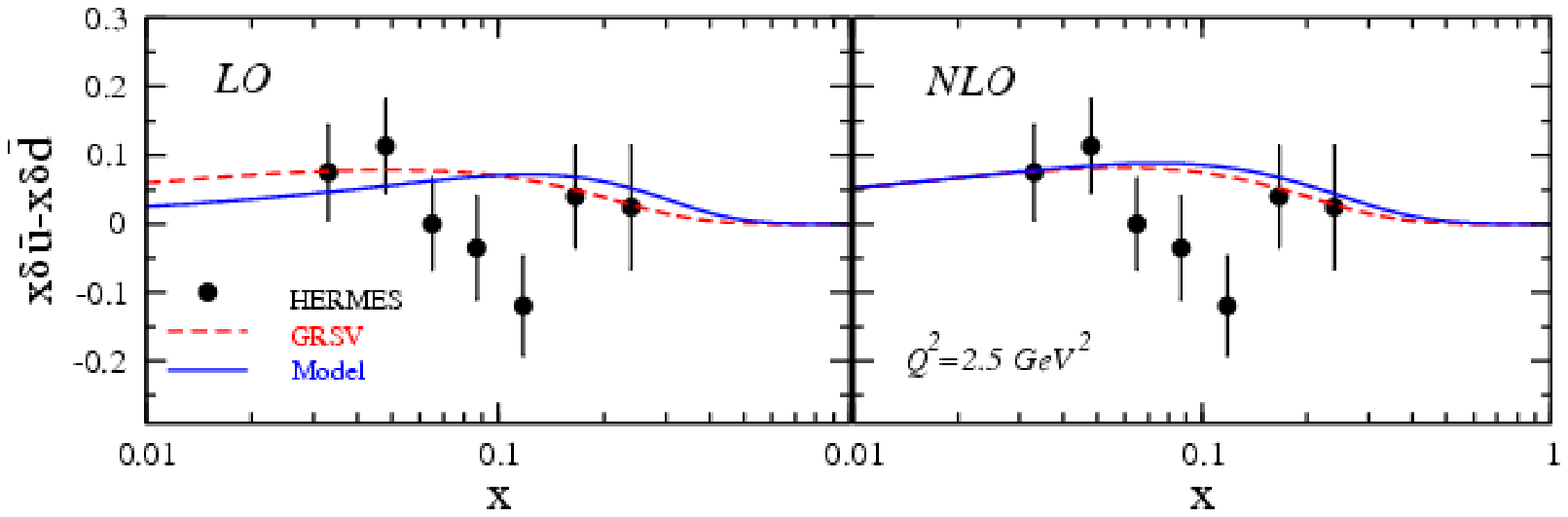}}
\caption{The flavor asymmetry in the helicity densities of the
light sea evaluated at $Q^2=2.5 GeV^2$. The  dashed line is the
GRSV model [15](ISET=2 for the NLO and ISET=4 for the LO
approximation) and the solid line is our model.}
\end{figure}

\section{Moments of helicity distributions}
In this section we study the first moment of the polarized parton
distributions in the broken scenario.  The total helicity of a
specific parton $f$ is given by the first ($n=1$) moment
\begin{equation}
 \label{eq:momentf}
\Delta f(Q^2) \equiv \int_0^1 dx\, \delta f(x,Q^2)\, .
\end{equation}\\
The contribution of various polarized partons in a valon are
calculable and by computing their first moment, the spin of the
proton can be computed. In the framework of QCD the spin of the
proton can be expressed in terms of the first moment of the total
quark and gluon helicity distributions and their orbital angular
momentum, i.e.
\begin{equation}
 \label{eq:SumRule}
 \frac{1}{2}= \frac{1}{2} \Delta\Sigma(Q^2) + \Delta g(Q^2) +
   L_{z}(Q^2)\;,
\end{equation}
which is called the helicity  sum rule. Here $L_{z}$ refers to the
total orbital contribution of all
(anti)quarks and gluons to the spin of the proton.\\

The contributions of $\delta \Sigma(x,Q^2)$ and $\delta g(x,Q^2)$
to the spin of proton can be calculated as in following.  The
singlet contributions form $j$-valon in a proton can be extracted
via \cite{ref17} \ba  \label{eq:delsig}\delta
\Sigma^j(x,Q^2)&=&\int_x^1 \delta f^{S}(\frac{x}{y},Q^2)[\delta
{\cal {W}}^{'}_{j}(y)\times G_{j/p}(y)]\frac{dy}{y}\;, \ea and the
gluon distribution for $j$-valon is \ba \label{eq:delglu}\delta
g^j(x,Q^2)&=&\int_x^1 \delta f^{gq}(\frac{x}{y},Q^2)[\delta {\cal
{W}}^{}_{j}(y)\times G_{j/p}(y)]\frac{dy}{y}\;, \ea where the
functional form for $\delta {\cal {W}}^{}_{j}(y)$ and $\delta
{\cal {W}}^{'}_{j}(y)$
have been defined in Subsec. 3.3.\\
Using Eq.~(\ref{eq:momentf}) and
Eqs.~(\ref{eq:delsig},\ref{eq:delglu}) we can arrive at the first
moments of the related parton distributions as follows  \ba \Delta
\Sigma^j(Q^2)&=&\int_0^1 \delta \Sigma^j(x,Q^2) dx\;, \ea \ba
\Delta g^j(Q^2)&=&\int_0^1 \delta g^j(x,Q^2) dx\;. \ea The
resulting total quark and gluon helicity for a $U$-valon is \be
 \frac{1}{2}\Delta \Sigma ^{U}(Q^2)+\Delta g^{U}(Q^2)\;,
\ee and for $D$-valon  \be
 \frac{1}{2}\Delta \Sigma ^{D}(Q^2)+\Delta g^{D}(Q^2)\;.
\ee
 Since  each proton involves 2 $U$-valons and one $D$-valon, the total quark
 and gluon helicity for the proton is
\be  \label{eq:valonSumRule}2\left(\frac{1}{2}\Delta \Sigma
^{U}(Q^2)+\Delta
 g^{U}(Q^2)\right)
+
 \frac{1}{2}\Delta \Sigma ^{D}(Q^2)+\Delta g^{D}(Q^2)=
 \frac{1}{2}\Delta \Sigma (Q^2)+\Delta g(Q^2)\;.
\ee The proton's spin is carried almost entirely by the total
helicities of quarks and gluons and  at $Q_0^2$ in the NLO
approximation is given by
\begin{equation}
\frac{1}{2}\Delta\Sigma +\Delta g
  \simeq 0.637,
\end{equation}
which is calculated in the NLO approximation at $Q_0^{2}$ value.
This amount is equal to the unbroken scenario result \cite{ref17},
and thus a negative orbital contribution $L_{z}(\simeq -0.1373$)
is required at the low input scales in order to comply with the
sum rule Eq.~(\ref{eq:SumRule}). It is intuitively appealing that
the non perturbative orbital (angular momentum) contribution to
the helicity sum rule Eq.~(\ref{eq:SumRule}) is noticeable because
of hard radiative effects which give rise to sizeable orbital
components due to the increasing transverse momentum of
the partons.\\

By using the results of Sec. 3 for the polarized parton
distributions in broken scenario and the definition of first
moment for these distributions as defined in
Eq.~(\ref{eq:momentf}), the numerical results for the first
moments are calculable. Our NLO results are summarized in Table
${\bf I}$ at some typical values of $Q^2$.
\begin{center}
\begin{tabular}{|c||c|c|c|c|c|c|c|c|}
\hline $Q^{2}(GeV^{2})$ & $\Delta u_v$ & $\Delta d_v$ & $\Delta
\overline{u}$ & $\Delta \overline{d}$ & $\Delta s=\Delta
\overline{s}$ & $\Delta g$ & $\Delta \Sigma $ & $\Gamma _{1}^{p}$
\\ \hline
$1$ & 0.5760 & - 0.0280 & 0.0907 & -0.2139 & -0.0615 & 0.5480 & 0.1786 & 0.1214 \\
$2.5$ & 0.5731 & -0.0269 & 0.0901 & -0.2145 & -0.0621 & 0.7417 &
0.1732 & 0.1238
\\
$5$ & 0.5716 & -0.0264 & 0.0899 & -0.2148 & -0.0624 & 0.8737 & 0.1706 & 0.1249 \\
$10$ & 0.5706 & -0.0260 & 0.0897 & -0.2150 & -0.0626 & 0.9986 & 0.1688 & 0.1258 \\
\hline
\end{tabular}
\end{center}

\textbf{Table I} First moments (total polarizations) of polarized
parton densities and $g_1^{p}(x,Q^2)$, as defined in
Eq.~(\ref{eq:momentf}) and in the flavor-broken scenario.
\\

\section{Orbital angular momentum}
The fundamental program in high energy spin physics focuses on the
spin structure of the nucleon. The nucleon spin can be decomposed
conceptually into the spin of its constituents according to
helicity sum rule. In last section we analyzed $\Delta\Sigma$ and
$\Delta g$ for each $Q^2$ value from polarized valon model. Here
we calculate the total orbital angular momenta of quarks and
gluons, $L_z(Q^2)\equiv L_{z}^q(Q^2)+L_{z}^G(Q^2)$. We know that
two places where the orbital angular momentum plays a role. One is
the compensation of the growth of $\Delta G$ with $Q^2$ by the
angular momentum of the quark-gluon pair. The other is the
reduction of the total spin component $\Delta\Sigma$ due to the
presence of the quark transverse momentum in the lower component
of the Dirac
spinor which is traded with the quark orbital angular momentum. \\

The evolution of the quark and gluon orbital angular momenta was
first discussed by Ratcliffe \cite{ref36}. A complete leading-log
evolution equation have been derived by  Ji, Tang and Hoodbhoy
\cite{ref32}:
\begin{eqnarray}
\frac{d}{dt}\left(
\begin{array}{l}
L_{z}^{q} \\
L_{z}^{G}
\end{array}
\right) =\frac{\alpha _{s}(t)}{2\pi }\left(
\begin{array}{ll}
-\frac{4}{3}C_{F} & \frac{n_{f}}{3} \\
\frac{4}{3}C_{F} & -\frac{n_{f}}{3}
\end{array}
\right) \left(
\begin{array}{l}
L_{z}^{q} \\
L_{z}^{G}
\end{array}
\right) +\frac{\alpha _{s}(t)}{2\pi }\left(
\begin{array}{ll}
-\frac{2}{3}C_{F} & \frac{n_{f}}{3} \\
-\frac{5}{6}C_{F} & -\frac{11}{2}
\end{array}
\right) \left(
\begin{array}{l}
\Delta \Sigma  \\
\Delta G
\end{array}
\right) , \end{eqnarray}
 with the solutions
\begin{eqnarray} \label{eq:Lzq} && L_z^q(Q^2) =
-{1\over 2}\Delta\Sigma+{1\over 2}\,{3n_f\over 16+3n_f}+
f(Q^2)\left(L_z^q(Q^2_0)+{1\over 2}\Delta\Sigma
-{1\over 2}\,{3n_f\over 16+3n_f}\right),\\
 \label{eq:Lzg} && L_z^G(Q^2) = -\Delta G(Q^2)+{1\over 2}\,{16\over 16+3n_f}+
f(Q^2)\left(L_z^G(Q^2_0)+\Delta G(Q^2_0)-{1\over 2}\,
{16\over 16+3n_f}\right), \nonumber \\
&& \end{eqnarray} where \begin{eqnarray} f(Q^2)=\left(
{\alpha(Q^2) \over \alpha(Q^2_0) } \right)^{{32+6n_f\over
33-2n_f}}\;,
 \end{eqnarray}
 and
$\Delta\Sigma$ is $Q^2$ independent to the leading-log
approximation. We see that the growth of $\Delta G$ with $Q^2$ is
compensated by the gluon orbital angular momentum, which also
increases like $\ln Q^2$ but with opposite sign. \\

If we know the total value of quark and gluon orbital angular
momentum at $Q^2_0$, then according to
Eqs.~(\ref{eq:Lzq},\ref{eq:Lzg}), $L_{z}^q(Q^2)+L_{z}^G(Q^2)$ can
be computed at each value of $Q^2$. By evaluating
Eq.~(\ref{eq:valonSumRule})  and using the sum rule,
Eq.~(\ref{eq:SumRule}) at $Q_0^2$, we can obtain the total quark
and gluon orbital angular momentum at $Q^2_0$. By adding
Eqs.~(\ref{eq:Lzq},\ref{eq:Lzg}) and inserting
$L_{z}^q(Q_0^2)+L_{z}^G(Q_0^2)$ in the additive equation, the
$Q^2$-evolution of total quark and gluon orbital angular momentum
will be accessible. The result for $L_{z}^q(Q^2)+L_{z}^G(Q^2)$ in
range of $1\leq
Q^2(GeV^2)\leq 1000$ has been depicted  in Fig. (7) .\\
\\
\begin{figure}[tbh]
\centerline{\includegraphics[width=0.55\textwidth]{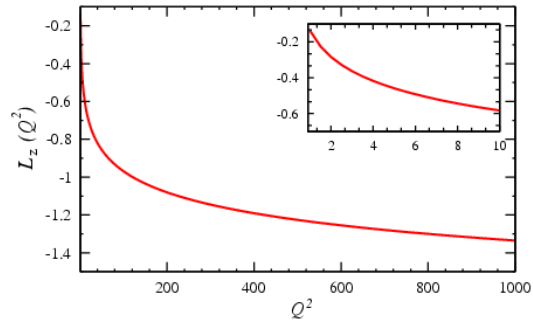}}
\caption{The $Q^2$ evolution of $L_z$ in the LO approximation. }
\end{figure}

On the other hand, we can directly obtain $L_z(Q^2)$, using the
helicity sum rule Eq.~(\ref{eq:SumRule}). In this case  the
$\frac{1}{2} \Delta\Sigma(Q^2)$ and $\Delta g(Q^2)$ are known from
polarized valon model in the LO and  NLO approximation. By having
different values of these two functions at some values of $Q^2$,
inserting them in the related sum rule and fitting over the
calculated data points for $L_z(Q^2)$, the functional form for
$L_z(Q^2)$ will be obtained in these two approximations.
The result for $L_z(Q^2)$ in the LO approximation, computed from
Eqs.~(\ref{eq:Lzq},\ref{eq:Lzg}), is completely consistent  with
the one from the sum rule in Eq.~(\ref{eq:SumRule}). If we use
from the data points for $L_z(Q^2)$, arising out from the
Eq.~(\ref{eq:SumRule}) but in the NLO approximation, we will
obtain the functional form $L_z(Q^2)=-5.10+4.96 (Q^2)^{-0.04}$.
Since the obtained results for $L_z(Q^2)$ in the LO approximation,
using evolution equations or fitting procedure, are in agreement
with each other, so it will be possible to say that in the NLO
approximation the  obtained fitting functional form for $L_z(Q^2)$
will be correspond to what will be resulted from evolution of this
quantity.
\section{Conclusions}
Polarized deep inelastic scattering (DIS) is a powerful tool for
the investigation of the nucleon spin structure. Experiments on
polarized deep inelastic lepton-nucleon scattering started in the
middle 70s. Measurements of cross section differences with the
longitudinally polarized lepton beam and nucleon target determine
the polarized nucleon structure functions $g_1(x, Q^2)$. During
the period of 1988-1993, theorists tried to resolve the proton
spin enigma and seek explanations for the  measurements of first
moment of proton structure function, $\Gamma^p_1$. These moments
were produced by European Muon Collaboration (EMC), assuming the
validity of the data at small $x$ and of the
extrapolation procedure to the unmeasured small $x$ region.\\

The determination of the polarized proton content of the nucleon
via measurements of the inclusive structure function $g_1(x, Q^2)$
dose not provide detailed information concerning the flavor
structure of these distributions. In particular the flavor
structure of the anti-quark (sea) distributions is not fixed and
one needs to resort to semi-inclusive deep in elastic hadron
production for this propose. The resulting anti-quark
distributions $\delta\bar q$ are, however, not reliably determined
 by this method  for the time being due to their dependence on the
 rather poorly known quark fragmentation function at low scales.
 Since now there are enough experimental date from semi-inclusive
 DIS experiments at DESY (HERMES), we followed the strategy of
 \cite{ref35} to break see quark distributions but in frame work of
 polarized valon model. The comparison of our obtained
 results for $\delta \bar {u}(x, Q^2),\delta  \bar {d}(x,
Q^2)$ and $\delta \bar {s}(x, Q^2)$ with the only available GRSV
model \cite{ref15}
 and experimental date from HERMES group indicate a very good
 agreement with them specially for strange sea quark while we
 expect to have a negative strange-quark polarization.
 Since the total contribution of sea quarks is remaining constant
 in two unbroken and broken scenario, the first moment of $g_1^p$,
 $\Gamma_1^p$, will be fixed, as is expected.\\

If we back  again to spin nucleon subject, we see that the
measurements by the EMC first indicated that only a small fraction
of the nucleon spin is due to the spin of the quarks\cite{ref19}.
Thus it refers to existence of other components in performing the
spin of nucleon. In fact the nucleon spin can be decomposed
conceptually into the angular momentum contributions of its
constituents according to the Eq.~(\ref{eq:SumRule}) where the
rest two terms of this equation give the contributions to the
nucleon spin from the helicity distributions of the quark and
gluon respectively. The angular momentum contribution has been
calculated in the LO approximation, using its $Q^2$ evolution in
Eqs.~(\ref{eq:Lzq},\ref{eq:Lzg}). If we know the parton
distributions in the NLO approximation, which obviously are known
using polarized valon model \cite{ref17}, then it will possible to
calculate the $\Delta \Sigma$ and $\Delta g$ contributions and
finally using the
helicity sum rule, to compute the $L_{z}$ contribution.\\

Extracting the  $L_z^q(Q^2)$ and $L_z^g(Q^2)$ which refer to
separate contributions of quark and gluon angular momentum, using
the polarized valon model, would also be valuable and challenging.
We
hope to report  on this subject in further publications.\\

The numerical data for the polarized quark distributions in
unbroken and broken scenarios are available by electronic mail
from $Alinaghi.Khorramian@cern.ch$.

\section{Acknowledgments}
We are thankful to R. C. Hwa for his useful comments to help us in
constructing the unbroken polarized valon model. The authors are
indebted to M. Mangano to read the manuscript. We acknowledge the
Institute for Studies in Theoretical Physics and Mathematics (IPM)
for financially supporting this project.

\end{document}